\documentclass[aps, prl, reprint]{revtex4-2}

\usepackage{graphicx}% Include figure files
\usepackage{dcolumn}% Align table columns on decimal point
\usepackage{bm}% bold math
\usepackage{caption}
\usepackage{hyperref}
\usepackage{amssymb}
\usepackage{float}
\usepackage{comment}
\usepackage{caption}
\begin{document}

\title{A guided light system for agile individual addressing of Ba$^+$ qubits with $10^{-4}$ level intensity crosstalk}

\author{Ali Binai-Motlagh, Matthew Day, Nikolay Videnov, Noah Greenberg, Crystal Senko, and Rajibul Islam}
\address{Institute for Quantum Computing and Department of Physics and Astronomy, University of Waterloo, Waterloo, Ontario, N2L 3G1, Canada}

\begin{abstract}
Trapped ions are one of the leading platforms for quantum information processing, exhibiting the highest gate and measurement fidelities of all contending hardware. 
In order to realize a universal quantum computer with trapped ions, independent and parallel control over the  state of each qubit is necessary. 
The manipulation of individual qubit states in an ion chain via stimulated Raman transitions generally requires light focused on individual ions. 
In this manuscript, we present a novel, guided-light individual addressing system for hyperfine Ba$^+$ qubits. 
The system takes advantage of laser-written waveguide technology, enabled by the atomic structure of Ba$^+$, allowing the use of visible light to drive Raman transitions. 
Such waveguides define the spatial mode of light, suppressing aberrations that would have otherwise accumulated in a free-space optics set up.
As a result, we demonstrate a nearest neighbour relative intensity crosstalk on the order of 10$^{-4}$, without any active aberration compensation. 
This is comparable to or better than other previous demonstrations of individual addressing.
At the same time, our modular approach provides independent and agile control over the amplitude, frequency, and phase of each channel; combining the strengths of previous implementations.  
\end{abstract}

\maketitle

\section{Introduction}

Trapped ions have the longest coherence times \cite{Harty2014, Wang2021}, highest fidelity single and two qubit gate operations \cite{BeBestFid, Ballance2016} as well as state-preparation and measurement (SPAM) fidelities \cite{Christensen_2020, Ransford2021,Fangzhao2022} of any experimental quantum information processing (QIP) platform. In recent years, barium has emerged as a popular candidate for trapped ion QIP due to the availability of a spin-1/2 isotope, long lived meta-stable states and visible wavelength atomic transitions. This favourable atomic structure has allowed for the highest experimentally demonstrated SPAM fidelities of any experimental qubit \cite{ba137SPAM}. For applications in QIP, the ability to manipulate the quantum state of individual ions is of utmost importance. Independent, coherent control of one or more qubits encoded in the hyperfine structure of an ion can be accomplished via stimulated Raman transitions. This experimental challenge is realized by tightly focusing individual laser beams at chosen ion sites across a chain \cite{Debnath2016, Nam2020}. The focus of this manuscript is on the necessary optical infrastructure needed for the individual addressing of long chains ($N > 10$) of Ba$^+$ ions with low crosstalk and independent control. The visible wavelength Raman transition (532 nm) in Ba$^+$ ions enables the use of laser-written waveguides as well as fiber coupled AOMs. The operation of the former has not been demonstrated in the UV,  while the latter is not commercially available for such short wavelengths. The ion of choice for many previous demonstrations of quantum computation has been $^{171}\mathrm{Yb}^+$, with 355 nm Raman transitions. $^{133}\mathrm{Ba}^+$ has all the same desirable properties of $^{171}\mathrm{Yb}^+$, with the added benefit of visible wavelength transitions, that enable the use of new optical technologies for robust quantum control of the ion. The ability to split and modulate light in waveguides, afforded by the use of such technologies, is central to the benefits of our implementation.

%\summary{Previous demonstrations are lacking in some way}
For complete quantum control over a chain of ions, an individual addressing system must provide agile (with a time scale faster than the single qubit gate time) and independent control over the temporal characteristics (intensity, frequency, phase) of each beam. 
These are key requirements for enabling arbitrary single qubit, and high fidelity multi-qubit entangling gates as well as several quantum simulation protocols \cite{FreqMod, Teoh2020}. Another major concern for the design of such a system is the intensity crosstalk between ion sites due to overlap of neighbouring laser beams. This presents a demanding optical engineering challenge as for most ions, the diffraction limited spot of the individual addressing beam is on order of a micron while the ion spacing is at most several microns in a typical chain. Thus, one must consider the error introduced to neighbouring sites when a given ion is addressed. Since this error is unitary, algorithmic techniques can be used to reduce the effects of crosstalk at the cost of greater circuit depth \cite{ParradoRodriguez2021crosstalk}. However, to get the most out of NISQ devices it is more favourable to reduce this error through optical engineering. A good target for the crosstalk intensity, relative to the intensity of the target ion, is 10$^{-4}$. Under a suitable optical scheme, this translates to a gate error on the order of $10^{-4}$ which is below the threshold for many error correcting codes for fault-tolerant quantum computation \cite{Benhelm2008}.

Previous demonstrations of individual addressing have been enabled by three predominant technologies: micro-mirrors, multi-channel acousto optic modulators (MAOMs) and acousto-optic deflectors (AODs). Table \ref{table:Techs} provides a comparison of each of these technologies. Beam steering with a single micro-mirror device allows the deflection of a high quality beam between array sites, leading to low crosstalk between sites at the expense of serial addressing \cite{crain2014individual}. Holographic beam shaping has further extended the micro-mirror device concept by using micro-mirror arrays, with each mirror much smaller than the beam size such that arbitrary beam profiles can be generated at ion sites \cite{Shih2021}. This allows for parallel, selective addressing with low crosstalk at the sacrifice of independent control of the frequency of each beam. While independent control over the intensity of each channel is possible, the slow switching rate of the micro-mirrors inhibits pulse shaping for the purpose of optimal control. AODs allow for agile intensity control, however, they also cannot provide independent frequency control at specific ion sites. On the other hand, independent and agile control of frequency, intensity  and phase can be accomplished through the use of MAOMs, which can generate an array of beams using a single diffractive optical element. However, since the MAOM is formed from a single crystal with individual acoustic transducers utilized to generate each beam, there is significant crosstalk between neighboring channels \cite{Debnath2016, Egan2021}. To combine full independent control of each beam with low cross talk, we propose a guided-light individual addressing system (GLIAS) that makes use of a laser-written waveguide splitter, manufactured using femtosecond laser direct-write (FLDW), and fiber coupled AOMs for the individual addressing of trapped ions. 
The use of waveguides suppresses aberrations that would otherwise be introduced in a free-space optical system.
Fiber coupled AOMs allow for Agile and independent control of each channel's temporal characteristics without distortion of the beam's spatial profile. 

\begin{table*}
\begin{tabular}{l|l|l|l|l|}
\cline{2-5}
&Intensity Crosstalk & Intensity Control & Frequency Control & Phase Control \\ \hline
\multicolumn{1}{|l|}{AOD \cite{Pogorelov2021}}   & 5e-3                 & \checkmark             &                  & \checkmark             \\ \hline
\multicolumn{1}{|l|}{DMD \cite{Shih2021}}   & 1e-4                 & \checkmark*                 &                   & \checkmark*             \\ \hline
\multicolumn{1}{|l|}{MCAOM \cite{Egan2021}} & 1e-3                 & \checkmark             & \checkmark           & \checkmark         \\ \hline
\multicolumn{1}{|l|}{GLIAS} & 1e-4                 & \checkmark              & \checkmark            & \checkmark        \\ \hline
\end{tabular}
\caption{Comparison of the characteristics of several Individual addressing systems. Check mark indicates that independent control (i.e. the ability to control a characteristic of one channel/beam without affecting that of the other channels) is possible. Asterisks in the DMD row indicate that independent control is possible, however it is slow relative to typical single qubit gate times.}
\label{table:Techs}
\end{table*}
%Similarly, it has been shown that a single or a pair of acousto-optic deflectors (AODs) can also be used for this purpose as well, where multiple RF frequency tones can be used to generate the individual addressing beams \cite{Pogorelov2021}. This scheme also exhibits low crosstalk at the expense of agile frequency control.

\begin{figure*}[htb]
    \centering
    \includegraphics[width=0.8\linewidth]{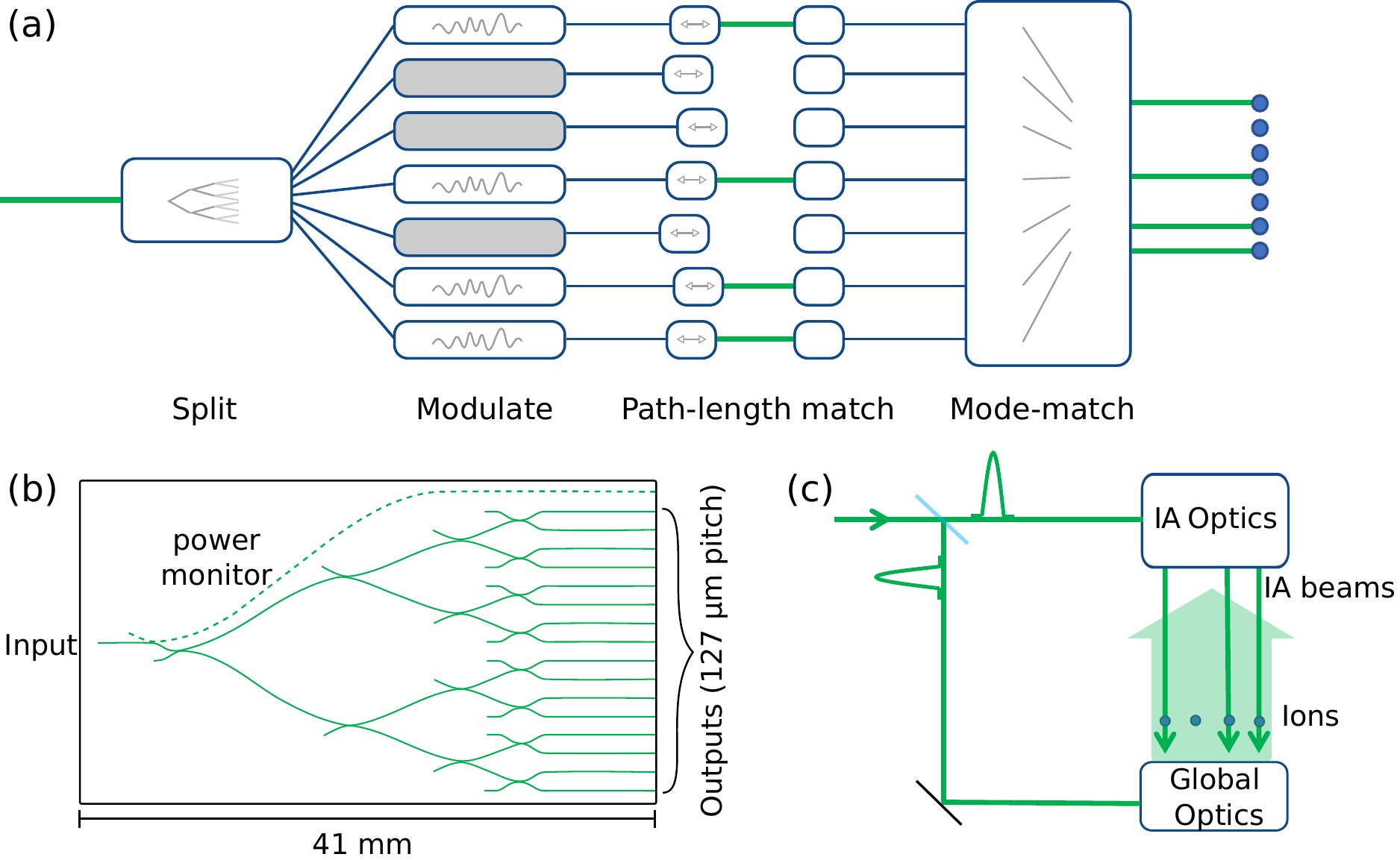}
\caption{(a) Schematic of the GLIAS, starting with the splitting done by the laser-written waveguide, modulation of the frequency and intensity of the light with fiber AOMs, temporal overlapping of the individual beams with delay stages, and spacial mode-matching of the ion chain with a micro-lens array before re-imaging onto the ion plane. (b) Diagram of the laser-written waveguide, the key enabling technology for the proposed GLIAS for use with Ba$^+$ ions. The laser-written waveguide is used to split light into 16 addressing channels, which can then each be fiber coupled, allowing independent modulation with fiber based AOMs. (c) Typical beam orientation for Raman transitions used to drive phase-stable single and two qubit gates in trapped ion chains \cite{lee}. The overlap of two beams is necessary to modify the quantum state of the ions. In this case one of the beams illuminates the entire chain while the other is tightly focused on each ions. It is also possible to replace this global beam with a second set of individual addressing beams. This can help reduce gate errors at the cost of additional engineering complexity.}
%The system is designed to couple to the normal modes along the radial direction of thEe ions in a Phoenix Trap provided by Sandia National Laboratories \cite{sandia}.
    \label{fig:Overview}
\end{figure*}

%%

%\summary{We present a system that doesn't lack in the requirements we have specified}

\begin{comment}
In this paper we present the design and demonstration of an individual addressing system that provides agile control of individual beams while maintaining low crosstalk between ion sites. The key enabling technology is a laser-written waveguide used to couple and split a single laser beam into multiple fiber channels. Once in fiber, standard fiber-optic coupled AOMs provide independent amplitude and frequency control before being routed through a final focusing telescope. The mode-quality of single-mode fiber ensures low crosstalk beams at the focal plane, given a well designed telescope. %We present the design of the proposed individual addressing system and then characterise a 16 channel guided-light individual addressing system designed for operation at a wavelength of 532 nm.
\end{comment}

An overview of this optical system and the orientation of the beams relative to the ions is shown in Fig.~\ref{fig:Overview}. The source is a 532 nm, mode-locked laser (NKT Photonics aeroPULSE PS) that gets divided into two beams. In the most common configuration, the profile in one arm is for global addressing and is shaped such that it illuminates the entire chain. The other arm is for individual addressing and is sent through a series of optics that splits the single beam into multiple channels and focuses each to a size significantly smaller than the ion spacing. The amplitude of each channel can be modulated with fiber AOMs to set the Rabi rate (the characteristic oscillation frequency of each qubit, that sets the single qubit gate time) for single qubit gates or to determine which ions are involved in a multi-qubit gate \cite{lee}. The normal modes of the ions are used as a vibrational data bus to create entanglement between two ions and hence enable the implementation of multi-qubit gates. Control over the applied frequency provided by each AOM determines which normal modes are used to mediate entanglement. This gives independent, agile control of each beam at the ion plane. We discuss each element of our proposed system, before presenting the results on the characterization of this individual addressing scheme.
%The two arms are counter propagating to maximize the momentum imparted by the beams onto the ions during entangling gates.
%%The AOMs can be used for the rapid modulation of the frequency and phase as well, which are used to optimize the fidelity of these operations

\section{System design}

The GLIAS is composed of four principal sections shown in Fig.~\ref{fig:Overview} (a): splitting, modulation, path-length matching and mode-matching. Each will be discussed in detail in the following sections.

\subsection{Splitting and modulation}
The splitting of a single beam from the laser source is implemented via a custom laser-written waveguide array (manufactured by OptoFab, Access Macquarie Ltd), which is written inside a monolithic block of alumino-borosilicate glass shown in Fig.~\ref{fig:Overview} (b). The chip takes a single beam and sequentially splits the light through concatenated 50/50 evanescent directional couplers. The initial input beam is split into 16 channels for the individual addressing of ions. The amount of coupled light between two waveguides is determined by the proximity as well as length of the waveguides and typical spacing is on the order of microns. The waveguides are created by using a high power, tightly focused femtosecond laser that locally heats up the piece of glass onto which the waveguide is being written. This locally changes the bonding structure of the underlying material at the focus, which increases the local refractive index by a small amount $\approx 10^{-3}$ relative to the surrounding material \cite{LDW}.

Power coupling from free space into the waveguide is done with a 3-axis translation stage and an objective (20 mm EFL, 15 mm WD, 0.25 NA). The input waveguide is recessed into the glass by 200 $\mu$m to increase the amount of power that can be incident on the input facet without damage. The input waveguide has a mode field diameter of $\sim$ 4.5 $\mu$m and 2 W of average power can be coupled without damage to the input facet. The waveguide array is coupled into a fiber array, where the 16 individual addressing outputs of the waveguide are each coupled to a single-mode polarization maintaining fiber (PM460). A power tap from the input waveguide is coupled into a multi-mode fiber. The waveguide and fiber array are bonded together using UV curable epoxy. Care was taken to ensure that epoxy does not enter the gap between the fiber and waveguide as the epoxy can be damaged from the high optical power, thereby reducing the transmittance of the device. The entire device is placed inside an aluminum case that serves to protect and strain relieve the fragile waveguide-fiber bond.

Each of the 16 channels is sent to a fiber coupled AOM (manufactured by Gooch $\&$ Housego PLC). The AOMs act as fast optical switches that can be used to precisely time when light is incident on a given ion in a chain, an essential requirement for individual quantum control. These AOMs also allow independent modulation of the amplitude, frequency and phase of each channel, which are key for complete control over the quantum state of the entire chain.
\begin{comment}
Quantum computing with trapped ions is based on spin-spin interactions, which is mediated by the motional modes of the ions. For a large chain of ions, there are many motional modes that each act as a ``vibrational data bus" between spins, transmitting information through multiple paths. This necessitates unique modulation of the amplitude or frequency of the light of each channel sent to individual ions. Optimization of the amplitude and frequency with each AOM ensures that no residual spin-motion entanglement remains at the end of the laser pulse \cite{Roos_2008}, allowing the native Mølmer–Sørensen gate to be used as a building block for high-fidelity logical gates \cite{Figgatt2018}. Beyond quantum computation, the all to all connectivity afforded by the Coulomb interaction allows trapped ion chains to be used for the simulation of arbitrary spin models. This again requires independent and parallel control over the frequency, amplitude and phase of the laser beams focused down onto the ions. These obstacles can be overcome when dealing with long chains of ions by first splitting the light with a laser-written waveguide and modulating the RF signal sent to the subsequent fiber-coupled AOMs. 
\end{comment}

\subsection{Path length matching}

To drive stimulated Raman transitions, the counter propagating beams must have spatial and temporal overlap. The spatial overlap can be obtained by suitable imaging techniques or monitoring ion signals that rely on IA Raman beams (for example measuring the AC stark shift).  Temporal overlap is made difficult as the fiber cables length tolerances are on the order of several millimeters. As the pulses from the 532 nm mode-locked laser are 10 picosecond in duration, corresponding to an approximately $L = $ 3 mm pulse length, the paths must be matched to a length $\Delta L \ll L$. To temporally align all pulses, miniature fiber delay stages with 4 mm of travel and a resolution of 210 $\mu$m (manufactured by OZ Optics Ltd.) are used for each individual addressing channel. For channels that have a length mismatch beyond what can be accommodated by the 4 mm travel range, the corresponding fibers can be cleaved and spliced.
\begin{figure*}[htb]
    \centering
    \includegraphics[width=0.8\linewidth]{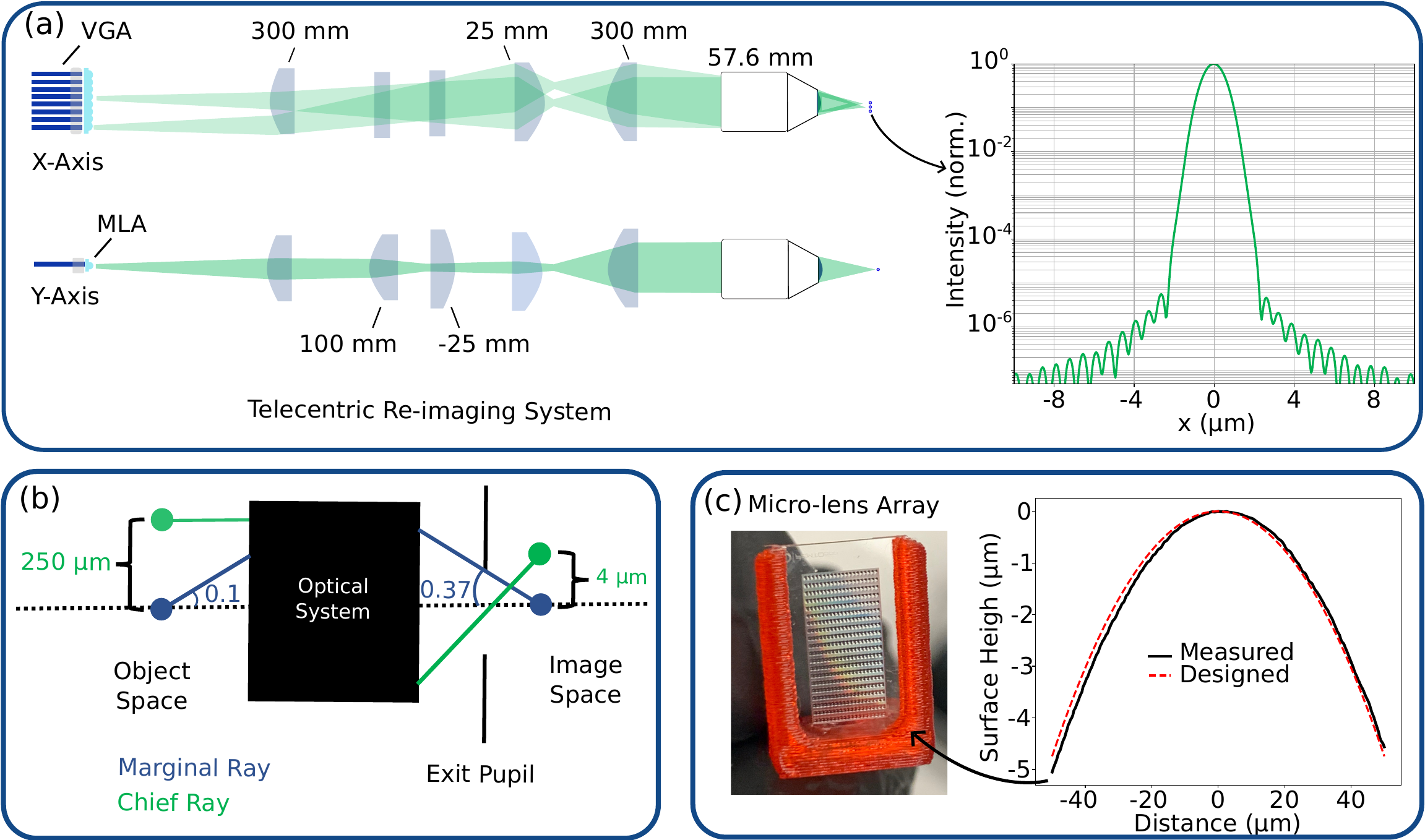}

    \caption{(a) Telecentric re-imaging system for mapping the microlens array (MLA) image onto the ion plane. The beam shape at the ion is elliptical, with a size orthogonal to the chain that is 4 times larger than that along the chain. This was done to reduce the magnitude of intensity fluctuations on the ion due to instabilities in beam pointing. The elliptical beam requires separate reshaping of the beam waists along the x-axis and y-axis with cylindrical lenses before focusing at the ion plane with an acromatic objective (manufactured by Special Optics Inc.). OpticStudio simulations of the focus at the ion plane through the MLA-Telescope optical system show $< 10^{-5}$ crosstalk along the ion chain axis (x-axis). (b) Optical re-imaging constraints set by the vacuum viewport (0.37 NA) \cite{noah} and ion spacing (4 $\mu$m), necessitating a reduction in object NA by a MLA in order to satisfy crosstalk requirements for high fidelity gates \cite{nik}. (c) Profile and picture of the MLA used in part to match the spatial mode of the VGA to the calculated ion spacing. The lenses contained in each row have the same profile, but each row of lenses are different. A single row from the MLA was used to create an image at the ion plane. The MLA was also designed in OpticStudio.}
    \label{fig:MLADesign}
\end{figure*}
\subsection{Mode-matching}
The guided light system terminates at the output of a 16 channel fiber array with a 3.3 $\mu$m mode diameter and a 250 $\mu$m mode-spacing. This must be mapped onto the 4 $\mu$m spacing between the ions in the chain. To satisfy the pitch change, a telecentric imaging system with a demagnification of 62.5x was built, with a design shown in Fig. \ref{fig:MLADesign} (a). The pitch between the ions is fixed at 4 $\mu$m (chosen as it enables us to attain our desired crosstalk figure while still allowing for relatively high normal modes frequencies) and the maximum available NA is fixed at 0.37 by the geometry of the trap and the vacuum re-entrants. Given these two constraints at the ion plane, we cannot directly image every fiber core and still satisfy the Lagrange invariant \cite{greivenkamp_2004}. 

%The choice of ion spacing represents a tradeoff between two-qubit gate times and nearest neighbour intensity crosstalk. Larger spacing allows for reduced crosstalk at the expense of slower gates. In our system, 4 $\mu$m was chosen as it allowed us to attain our desired crosstalk figure while still allowing for relatively high normal modes frequencies. 

This problem is depicted in Fig. \ref{fig:MLADesign} (b). 
At conjugate planes, the product of the height of the chief ray and the angle of the marginal ray is conserved for paraxial optical systems (perfect imaging). 
There are two plausible solutions. One could reduce the pitch between channels, thus reducing the height of the Chief ray or one could reduce the fiber NA, thus reducing the angle of the Marginal ray.
The former approach was dismissed as simulations indicated significant crosstalk between neighbouring waveguides  \cite{nik}. To reduce the NA of each fiber core, an array of microlenses are placed at the facet of each fiber to expand the beam waist out of each channel. The custom microlenses were designed using OpticStudio and manufactured by FEMTOprint. An aspheric profile, optimized by considering the resulting crosstalk in the ion plane as well as typical alignment tolerances, was used to minimize spherical aberrations in the system. A picture of these microlenses is shown in Fig. \ref{fig:MLADesign} (c).

Simulations of the microlenses with the telecentric imaging system indicate a nearest neighbour intensity crosstalk of $< 10^{-5}$. The simulated beam profile after the telecentric imaging system is shown in Fig. \ref{fig:MLADesign} (a). The individual addressing system is designed to be telecentric so that the wavevectors of the beams from different channels are all parallel to each other and orthogonal to the chain. This is necessary to minimize coupling to the axial modes of the chain, which are not intended to be used for QIP in this optical configuration.

\section{Results}

\subsection{Splitting ratios and maximum power throughput}

\begin{figure*}[htb]
    \centering
    \includegraphics[width=0.8\linewidth]{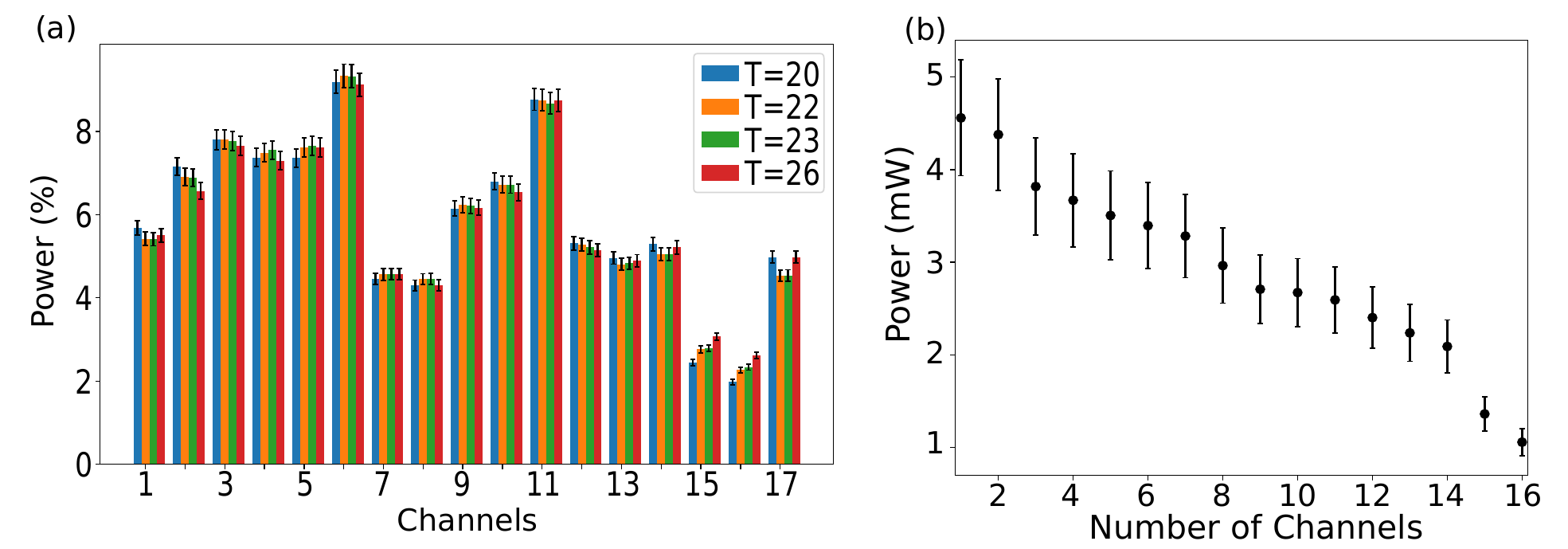}
    \caption{(a) Relative power output of a single-mode fiber glued to each channel of the laser-written waveguide. (b) The maximum balanced power per channel as a function of the number of channels included. Error bars are calculated based on the measurement uncertainty of the power sensor used for measurements ($\pm 3\%$).}
    \label{fig:Splits}
\end{figure*}

All channels coming out of the waveguide were designed to have equal optical powers, so that each ion experiences the same Rabi rate. However, as shown in Fig. \ref{fig:Splits} (a) we observe relatively large, unpredictable variations in the power output of each channel, due to uncontrolled fabrication variations in individual couplers.  There is roughly a factor of 5 difference in the optical power out of the best performing channel (6) and the worst performing channel (16). The channel with the lowest optical power determines the maximum single qubit gate time of the computer. This variation adds a slight complexity to the experiment as the Rabi rate of each ion must be independently known and tracked. If equal Rabi frequencies are desired, the powers on each channel must be made equal, which can be done by controlling the RF power sent to the individual AOMs.

Knowledge of the power reaching the ions is of utmost importance as it sets an upper bound on the possible Rabi rate. Insertion losses for major components of the individual addressing system are shown in Table \ref{table:IL}. If equal optical power is desired for all channels, the input power must be further attenuated by the AOMs so that all channels have the same power as the channel with the smallest splitting ratio. This balanced output power can be increased by using a subset of all the available channels, ignoring one or more of the channels with the smallest output power. This is because through this omission, the AOMs must compensate for a smaller variation in the unbalanced power out of each channel. The curve in Fig. \ref{fig:Splits} (b) shows how the power per channel at the ions increases as the number of included channels are reduced, where each time we choose not to use the worst-performing channel in the set. To attain a Rabi rate of around $2\pi\times1~$MHz, we need the optical power to be around 2 mW on the individual addressing side, assuming equal optical intensities on both arms of the individual addressing system (see supplementary). From this curve it is evident that to attain our desired $2\pi\times1~$MHz Rabi rate, we must exclude the two worst-performing channels (channels 15 and 16). It may be possible to extract more power from the device by splicing each subsystem of the individual addressing device together and shortening the length of the fibers. This data is based on an input average power of 2 W into the waveguide chip, which is below the damage threshold for the chip. 

%The input power is further attenuated at the AOMs by the requirement that all channels must have equal optical powers. To attain a Rabi rate of around 1 MHz, we need the optical power to be around 1.5 mW on the individual addressing side. A Rabi rate around 1 MHz results in a $\pi$-pulse time of less than 1 $\mu$s. The red (black) curve in Fig. \ref{fig:Splits} (b) shows how the maximum obtainable Rabi frequency (power per channel at the ions) increases as the number of included channels are reduced, where each time we choose not to use the worst performing channel in the set. The power per channel increases each time a channel is excluded for individual addressing since the AOMs must compensate for a smaller variation in the powers out of each channel. From this curve it is evident that to attain our desired 1 MHz rabi rate, we must abandon the two worst performing channels (channels 15 and 16). With this omission we can extract more than 1.5 mW from each channel. It may be possible to extract more power from the device by splicing each subsystem of the individual addressing device together and shortening the length of the fibers. This data is based on an input average power of 2 W into the waveguide chip. 
%Excluding channels 15 and 16, the total insertion loss of the system is around -19 dB as indicated by the black curve in Fig. \ref{fig:Splits} (b).

To protect against power fluctuations due to temperature gradients produced by ambient fluctuations or scattered light, active control of the temperature was built into the protected waveguide casing. Varying the temperature from 20-26 $^\circ$C, well beyond the expected change in the temperature of the device, we observe a relatively small change in the splitting ratio (at most 0.6\%). This data is shown in Fig. \ref{fig:Splits} (a).

\begin{table}[H]
\centering
\begin{tabular}{l|l}
Element                     & \begin{tabular}[c]{@{}l@{}}Insertion loss (dB) \\\end{tabular}  \\ 
\hline
Waveguide                   & 5  $\pm 0.4$        \\
Waveguide-VGA bond + Fibers & 4.9 $\pm 0.4$  \\
AOM (single device) & 3 $\pm 0.4$                           \\
Variable Delay lines        & 1.9-4.6 $\pm 0.4$                    \\
VGA                         & 1.4  $\pm 0.4$                                                           \\
Telescope + Viewport        & 1.2  $\pm 0.4$   

\end{tabular}
\caption{Breakdown of the insertion loss of the major components of the guided-light individual addressing system. Besides loss of power from individual components, mismatch in the output power from different waveguide channels is a limiting factor in maximum obtainable uniform power at each ion site across the chain. Loss from power matching is characterised in Fig. \ref{fig:Splits}. Errors are calculated based on the measurement uncertainty of the power sensor used for measurements ($\pm 3\%$).}
\label{table:IL}
\end{table}

\subsection{Path length matching characterisation}
\begin{figure*}[htb]
    \centering
    \includegraphics[width=0.8\linewidth]{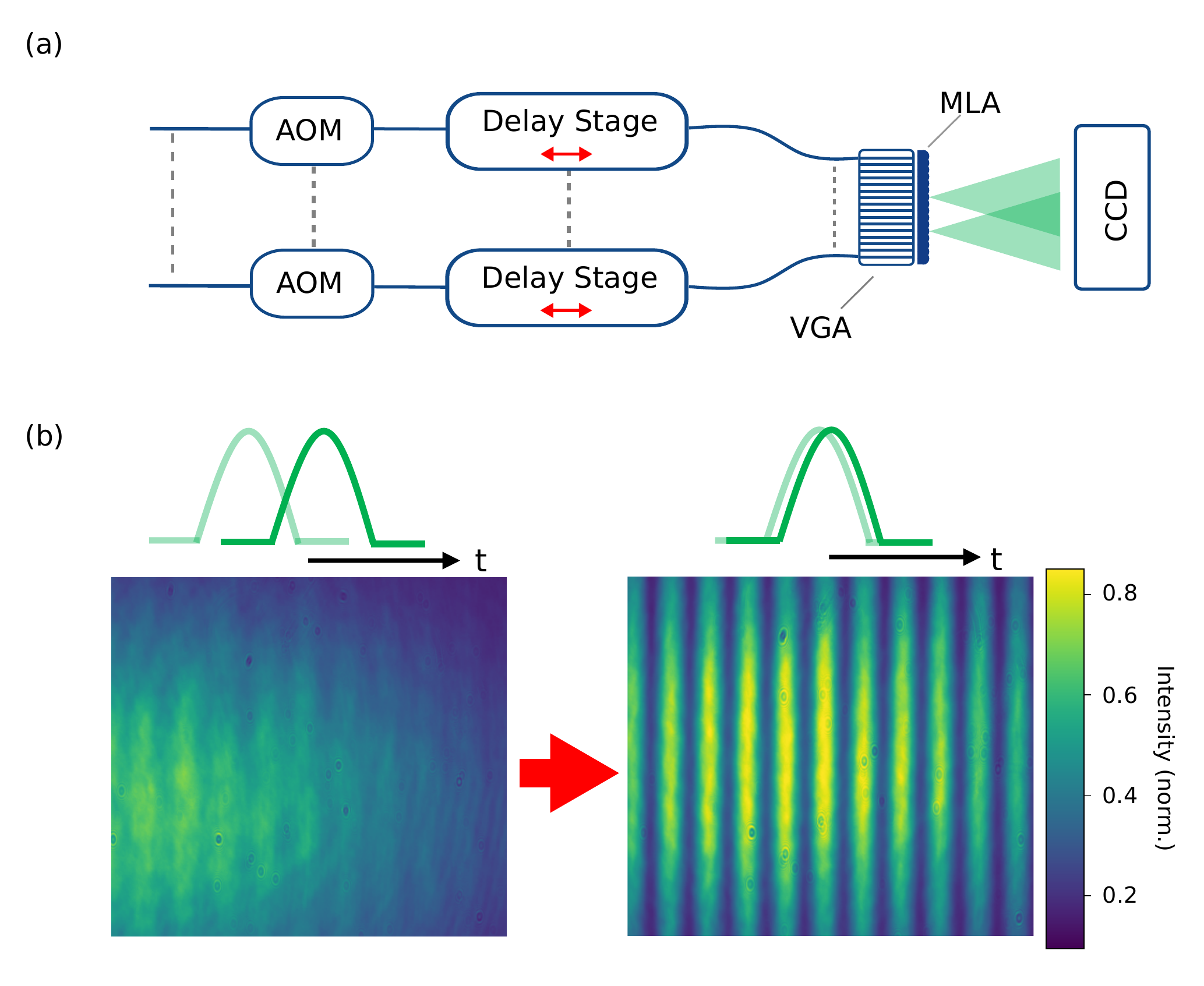}
    \caption{(a) Overview of the entire IA system with fiber delay stages to ensure that the pulses arrive at the ions at the same time. Stages with 4 mm travel range were used to achieve temporal overlap between a pair of beams in the individual addressing system. (b) Fringe contrast created by the overlap of the two beams. Maximizing the fringe contrast ensures that all pulses arrive at the ions at the same time. For the purpose of path length matching, all AOMs are set to the same frequency.}
    \label{fig:PathLength}
\end{figure*}

To optimize the temporal overlap we send light from two channels of the FLDW splitter chip through to the VGA and block the rest using the AOMs. The resulting interference pattern formed by the two beams is then captured using a camera, at a point where they have significant spatial overlap. If there is no temporal overlap between the pulses then no interference fringes will be detected on the camera. The overlap between the pulses can be optimized by maximizing the contrast of the resulting fringes, through displacement of the fiber delay stages. The experiment is described in Fig. \ref{fig:PathLength} (a) and the resulting fringe contrast is shown in Fig. \ref{fig:PathLength} (b). Prior to optimization, the fringes are faint and hard to detect. After optimization, we see a clear interference pattern with large fringe contrast. For some channels in the system, the 4 mm range of the stages was not sufficient to fully optimize the interference pattern. For these, we use a pair of stages with longer travel range to precisely determine the length mismatch then splice the fibers to correct for this difference. For further fine tuning of the spatial and temporal overlap, the measured Rabi frequency or AC stark-shift at the ions can be used as the feedback signal.

\subsection{Microlens array characterisation}
The microlenses are manufactured using ultrafast lasers to locally modify the index and density of a piece of glass. This technique combined with chemical etching can be used to realize complex three dimensional structures in glass, such as our microlenses. The initial set of fabricated microlenses possessed a shorter focal length than what we had designed for, possibly due to the lens polishing step in the manufacturing process. To account for these fabrication inaccuracies, a grid of microlenses was designed with a range of effective focal lengths (EFL) starting from 0.525 mm and going up to 1 mm in 0.025 mm increments. A total of 20 microlens arrays (MLA) were produced, with each array (row) containing identical microlenses. Through this process, we were able to identify a row with the desired EFL.

%To determine whether an array in this new batch of MLAs has the desired effective focal length, an optical profilometer (Bruker Contour 3D Optical Profiler) was used to measure the surface profile of one lens from each array. The profilometer is made up of a Michelson interferometer with a white light source. 
%Fig. \ref{fig:MLADesign} (c) Compares the measured curvature of a particular microlens in the grid and the curvature of the designed lens. On average, the fabricated lenses have an EFL that is about 2\% shorter than the design (see supplementary). Despite this discrepancy, by manufacturing lens arrays with several different EFLs, we were able to identify several rows with EFLs that result in low crosstalk at the ion plane.

\begin{figure*}[htb]
    \centering
    \includegraphics[width=0.8\linewidth]{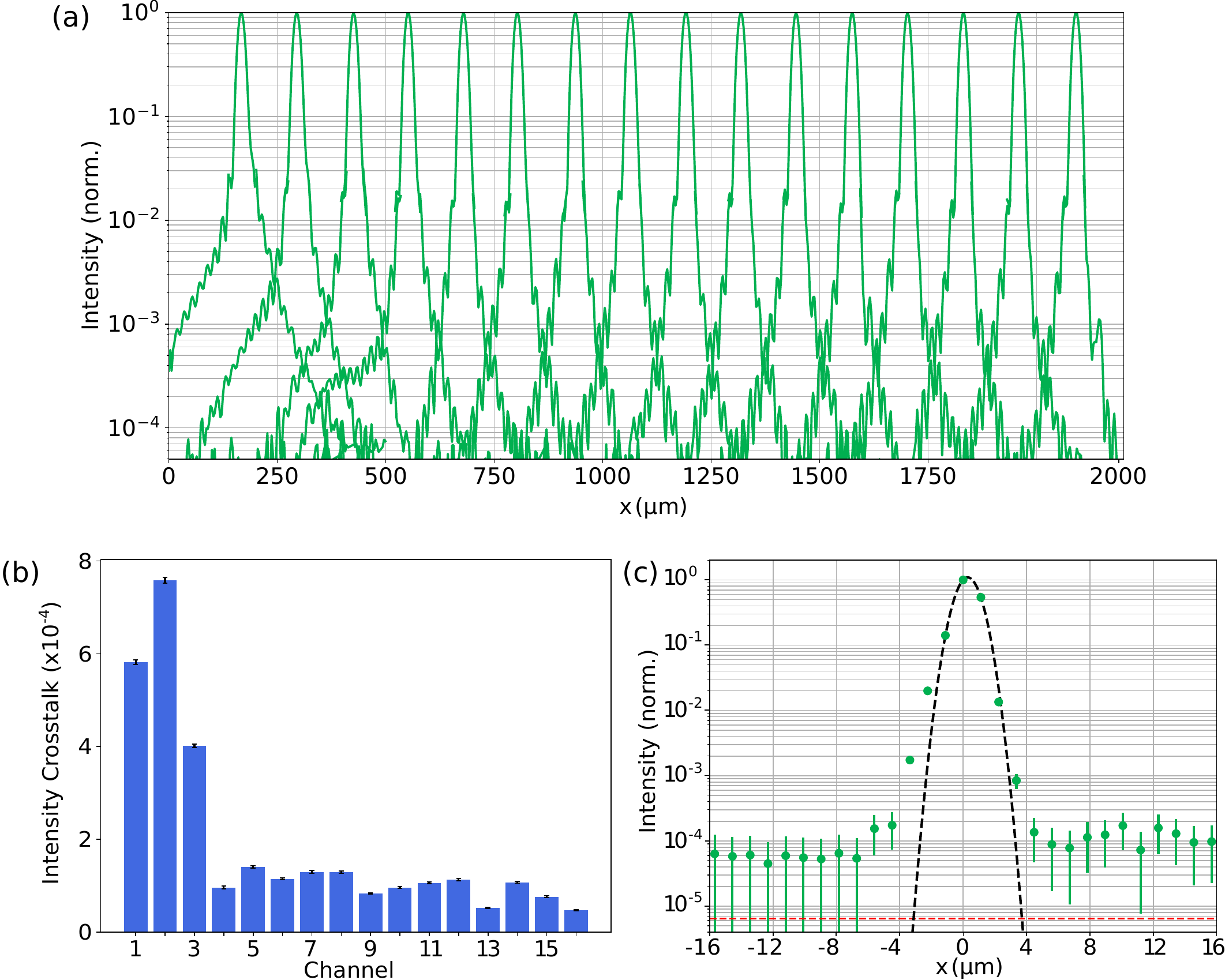}
    \caption{(a) High-dynamic range images of the IA beams measured on the DataRay WinCamD beam profiling camera. The beams after the MLA are imaged onto the camera using a single 250 mm focal length lens, providing a 0.5x magnification. Only two channels exhibit significantly higher crosstalk than expected, likely due to manufacturing imperfections of the MLA. Error bars are too small to be seen in the plot. (b) Intensity on each ion site when light is shining on its nearest neighbour(s). (c) Profile of a beam at the ion plane after the entire IA system, measured using the Raspberry Pi NoIR V2 camera. This shows crosstalk at the 10$^{-4}$ level, consistent with the prior measurements at lower magnification. The red dashed line indicates the noise floor for the measurement. Errors are calculated based on typical CMOS sensor dark noise, read-out noise as well as the photon shot noise.}
    \label{fig:CrossT}
\end{figure*}
\subsection{Crosstalk intensity characterisation}
To characterize the intensity crosstalk between channels, the beams after the MLA are imaged onto a camera using a single 250 mm focal length lens, providing a 0.5x magnification. Several profiles at multiple exposures were recorded and stitched together, to obtain a single profile with sufficient dynamic range to assess the crosstalk level between the neighbouring channels. The beam profile images are taken with one channel turned on at a time. The data for all 16 channels is shown in Fig. \ref{fig:CrossT} (a). The profiles near the left end of the the system are broadened compared to the rest of the channels and as a result exhibit a larger amount of intensity crosstalk. For the first three channels, the crosstalk is on the order of 10$^{-3}$. For the rest, it is on the order of 10$^{-4}$. This asymmetry is most likely due to imperfections in the MLA manufacturing process. Fig. \ref{fig:CrossT} (b) elucidates this by showing the crosstalk for each channel, when its nearest neighbours are simultaneously turned on. 

\begin{comment}
Better alignment may further reduce the intensity crosstalk, but the extinction ratio of fiber coupled AOMs are on the order of 10$^{-5}$, which is a more fundemental limit. 
\end{comment}
To verify these results, the crosstalk measurement was repeated for a single channel after the full individual addressing system (62.5x demagnification). To capture the beam profile at the ion plane, a camera with a relatively small pixel size is required, given that the expected beam waist at the focus of the objective is around 0.9 $\mu$m. For this we use the Raspberry Pi Noir V2 camera. It is a color camera with each unit cell consisting of 2 green pixels, one red pixel and one blue pixel. The size of each unit cell is 1.12 $\mu$m, thus the resolution is too low relative to the beam radius to obtain a precise beam profile. Nevertheless, the measured intensity crosstalk level should be comparable to that obtained in the prior measurement. This profile is shown in Fig. \ref{fig:CrossT} (c). Our measurements indicate an intensity crosstalk slightly greater than 10$^{-4}$, 4 $\mu$m away from the peak, at the location of the neighbouring ions. This is commensurate with the data shown in Fig. \ref{fig:CrossT} (a).

\section{Conclusion}
In this paper we demonstrated the design and characterization of a individual addressing system suited for digital quantum computation and analog simulation with a chain of up to 16 Barium ions. The system utilizes fiber optic and waveguide technology, making it modular and thus scalable and upgradable if addressing of larger chains is desired. This modularity also means that we could replace the waveguide chip, if we are able to design and manufacture one with a more balanced splitting ratio, without affecting the alignment of the system to the ions. Our architecture is not limited to FLDW technology. Future upgrades could replace this chip with higher index contrast, lithographically defined integrated waveguides, reducing crosstalk due to evanescent coupling. This crosstalk sets a threshold on the minimum possible waveguide pitch. Reduction of this threshold would allow the Lagrange invariant to be satisfied through a change in the pitch of the VGA fibers, thus removing the need for a MLA and thereby further reducing the number of mechanical degrees of freedom. This control would also allow for the pitch to be exactly tailored to the pitch of the ion chain which is necessary for chains with non-uniform spacing. 

We measure an intensity crosstalk on the order of 10$^{-4}$. If the other arm of the Raman system in Fig. \ref{fig:Overview} globally illuminates the ions then the resulting error in the Rabi rate is proportional to the square root of the intensity crosstalk. This corresponds to a crosstalk error in Rabi rate of 1\% \cite{Ozeri2007, Wineland2003}, which is comparable to the state-of-the-art. 
To make this more concrete, if a $\pi$-pulse is performed on a desired ion, this crosstalk means that its nearest neighbours undergo a $\pi$/100 pulse. 
To reduce this error further, one can implement individual addressing for both arms of the Raman system in which case the error in the Rabi rate scales linearly with the intensity (assuming equal intensity in the two beams).
This will result in a crosstalk error in Rabi rate of 0.01\%, commensurate with requirements of quantum error correction algorithms \cite{Benhelm2008}.
At the same time, our individual addressing system provides rapid, simultaneous control over the frequency, phase and amplitude of each beam using fiber coupled AOMs. 
This independent control is necessary for simulation of arbitrary fully connected spin models with ions \cite{Teoh2020, Rajabi2019}.
Further, the use of waveguides and fibers simplifies optical alignment and makes the system modular. Maintenance or upgrades to one part of the system has no effect on the alignment of components downstream from it. The use of this technology is not limited in scope to unitary qubit operations. FLDW waveguide technology is compatible with 493 nm, the S$_{1/2}$ to P$_{1/2}$ transition wavelength of the Barium ion, which is used for quantum state detection. Thus our system can be used for independent and agile state detection, in addition to unitary gate operations. This enables the implementation of mid-circuit measurement which is key to the realization of quantum error correction.
%This low level of cross talk, when implemented with a %resonant beam (for example 493 nm resonant transition) %will also allow high fidelity in-situ qubit measurement %and reset, further extending the capabilities of the %system for quantum error correction and new classes of %quantum simulation. Measurement-induced quantum phases %realized in a trapped-ion quantum computer

\section*{Acknowledgements}
This research was supported in part by the Natural Sciences and Engineering Research Council of Canada (NSERC), Grant Nos. RGPIN-2018-05253 and RGPIN-2018-05250, and the Canada First Research Excellence Fund (CFREF), Grant No. CFREF-2015-00011. CS is also supported by a Canada Research Chair.

\bibliographystyle{apsrev4-2}
\bibliography{references}
\end{document}